\documentclass[12pt]{article}
\usepackage{fancyheadings,times,amssymb,curvesls,harvard}
\newcommand{\Rset}{\mathbb{R}} 
\newtheorem{theorem}{Theorem}

\begin{document}

\begin{titlepage}
\begin{center}
{\Large{\bf Topology change and context dependence}}\\[5mm]
{\large Mark J Hadley}\\[10mm]
{\small Department of Physics, University of
Warwick, Coventry CV4~7AL, UK\\[5mm]
email: M.J.Hadley@warwick.ac.uk\\ Tel: +44 1203 522410\\ Fax:
+44 1203 692016}

\end{center}

\begin{abstract}
The non-classical features of quantum
mechanics are reproduced using models constructed with a classical
theory - general relativity. The inability to define complete initial
data consistently and independently of future measurements,
non-locality, and the non-Boolean logical structure are reproduced by
these examples.  The key feature of the models is the role of topology
change. It is the breakdown of causal structure associated with
topology change that leads to the apparently non-classical behaviour.
For geons, topology change is required to describe the interaction of
particles. It is therefore natural to regard topology change as an
essential part of the measurement process. This leads to models in
which the measurement imposes additional non-redundant boundary
conditions. The initial state cannot be described independently of the
measurement and there is a causal connection between the measurement
and the initial state.
\end{abstract}


\end{titlepage}

\section{Introduction}
It is well known that quantum mechanical systems cannot be described
as classical systems evolving in time independently of future
measurements\cite{belinfante,beltrametti_cassinelli}. Attempts to
construct classical models either fail or have non-local,
measurement-dependent, influences. The Kochen-Specker
paradox\cite{belinfante} is an example of the inability to assign
initial data (particle spins) independently of future
measurements. While in Bohm's theory particle positions and momenta
can be defined but they are insufficient to determine the subsequent
evolution - the desired results are only obtained at the expense of
introducing a non-local influence in the form of the pilot
wave. 

Common to any interpretation of quantum mechanics is the non-Boolean
logical structure\cite{isham95,beltrametti_cassinelli} of the
propositions. The propositions have to be defined differently in the
different interpretations (see~\cite[Section 9.2]{isham95}), so that
in some interpretations the propositions are about properties held by
the particle while in other interpretations this would be meaningless
and the propositions are statements about the state preparation. The
non-Boolean logic of propositions characterises quantum theory and
distinguishes it from classical theory. Once a measurement is defined,
and propositions are restricted to those that relate to the chosen
measurement apparatus, the results have the familiar logical structure
- that this must be the case is explained by Mackey
see~\cite[Chapter~13]{belinfante,beltrametti_cassinelli}. Unlike any
classical theory, quantum mechanics is therefore a context-dependent
theory.

It is known that if a breakdown in causal structure is an inherent
feature of elementary particles then the propositions associated with
the theory could have the, non-Boolean, logical structure of quantum
mechanics\cite{hadley97}. In this paper models are constructed upon
which it is impossible to define complete initial data independently
of future measurements. This context dependence is the essence of the
Kochen-Specker Paradox. Another striking characteristic of quantum
theory is the non-locality predicted, and confirmed, in EPR
experiments - this too is reproduced in these examples.

As early as 1957 Dennis Sciama pointed out that quantum mechanics was
a way to account for hidden variables where half of them were in the
future \cite{sciama} 

A common feature of the models presented in this paper, is a breakdown
of the causal structure in such a way that there exists a causal link
from the measurement to the {\em initial} data, and that the
measurement itself imposes non-redundant boundary conditions. It is
known that a theory with these properties will have the logical
structure of quantum mechanics. Apart from exotic
alternatives\cite{fivel}, any vector representation will have the
familiar Hilbert space structure, operators and commutation relations.

Models of elementary particles based on geons (where some or all of
the particle's properties arise from the topology of spacetime)
require topology change to occur when interactions take
place. Theorems of Geroch and Tipler place severe constraints on
topology change which prevent it occurring without a breakdown of the
causal structure in one sense or another. The relationship between
topology change and non-trivial causal structure is exploited in this
paper. By associating topology change with the measurement process in
quantum mechanics, we are naturally led to models which display
context dependence.

Geroch's theorem \cite{geroch} shows that for a compact time-oriented
manifold without closed timelike curves, CTCs, the topology cannot
change from one spacelike boundary to the other. Some authors consider
the causal structure to be more fundamental and have examined
spacetimes with singularities in order to allow topology change
\cite{dowker_garcia,sorkin97}. This paper does not consider
singularities, it retains the framework of classical general
relativity - including a continuous spacetime manifold with a
continuous metric. The breakdown in causal structure is actually seen
as advantageous because it offers a way of reconciling classical and
quantum physics \cite{hadley97}. The examples are created by relaxing
the assumptions of Geroch's theorem. We consider in turn manifolds
which are not time-orientable, those with CTCs, non-compact manifolds
and finally examples where the spacelike boundaries cannot be
defined. In each case the properties are compared with the phenomena
previously associated exclusively with quantum theories. If topology
change is considered to be an integral part of the measurement process
then these models are examples of context dependence which is the
distinguishing characteristic of quantum theory.

\section{Notation}
The paper is concerned with spacetime manifolds of 1+1, 2+1 and 3+1
dimensions with a Lorentzian metric of signature (-,+),(-,+,+) and
(-,+,+,+) respectively. The combination of the manifold and the
Lorentzian metric is referred to as a {\bf geometry}. The metric
defines a timelike ray (a pair of timelike vectors $\gamma$ and
$-\gamma$ corresponding to a forward and backward direction in
time). A geometry which is time-orientable admits a continuous choice
of timelike vector $\gamma$ and is called {\bf isochronous}. An
isochronous manifold $M$ therefore admits a continuous vector
field. Indeed a manifold which admits a continuous vector field can be
endowed with a Lorentzian metric such that it is isochronous see for
example~\cite{borde,reinhart}.

The constructions which follow comprise a geometry $M$ (of dimension
$n+1$) whose boundary is the disjoint union of two manifolds
$\Sigma_1$ and $\Sigma_2$ (each of dimension $n$). By a topology
change we mean that the topology of $\Sigma_1$ differs from that of
$\Sigma_2$. From the results above, a timelike vector field can be
constructed in the isochronous case. A unique timelike curve can be
constructed through any point such that the curve is everywhere
tangent to the vector field. The curves, but not the vector field, can
also be constructed in a geometry which is not isochronous.

Most of the paper is concerned with compact geometries. An important
class of non-compact spacetimes to which Geroch's theorem, and related
considerations, also apply are externally simple spacetimes (see Borde
\cite{borde}). A spacetime $M$ is externally simple if there exist
compact regions $\tilde\Sigma_1$ and $\tilde\Sigma_2$ such that
$\Sigma_1 - \tilde\Sigma_1$ is diffeomorphic to $\Sigma_2 -
\tilde\Sigma_2$.

\section{Theorems}
The following theorem, attributed to C W Misner, is a result special
to $3+1$ dimensions:

\begin{theorem}[Misner]
Let $\Sigma_1$ and $\Sigma_2$ be two compact 3 manifolds. Then there
exists a compact geometry $M$ whose boundary is the disjoint union of
$\Sigma_1$ and $\Sigma_2$, and in which $\Sigma_1$ and $\Sigma_2$ are
both spacelike.
\end{theorem}
A proof can be found in \cite{geroch,yodzis}. The significance of the
theorem is that, in the three space and one time dimension of primary
interest, a timelike vector field can always be constructed. There are
no topological obstructions as there are in other dimensions.

\begin{theorem}[Geroch]
Let $M$ be a compact 4-manifold with metric signature \mbox{(-,+,+,+)}
whose boundary is the disjoint union of two compact spacelike
manifolds, $\Sigma_1$ and $\Sigma_2$. Suppose $M$ is isochronous and
has no closed timelike curves (CTCs). then $\Sigma_1$ and $\Sigma_2$
are diffeomorphic and further, $M$ is topologically $\Sigma_1 \times
[0,1]$.
\end{theorem}
Although a 4-manifold was specified (three space and one time
dimension) in the original theorem, the proof is also valid for 1+1
and 2+1 dimensions.

Essentially, in order for topology change to take place, from one
spacelike boundary to another, at least one timelike ray must pass
through $\Sigma_1$ but not $\Sigma_2$ or vice versa. Geroch shows that
for compact manifolds there are only two ways for this to happen:
\begin{enumerate}
\item The timelike rays that cross the surface $\Sigma_1$ and enter
$M$, never leave the interior of $M$ - in which case, due to the
assumed compactness, there must be a CTC within $M$.
\item The timelike rays that cross the surface $\Sigma_1$ and enter
$M$, also exit through $\Sigma_1$ - which implies that $M$ cannot be
isochronous.
\end{enumerate}

The first case is further refined by Tipler, who assumes that $M$ is
isochronous:
\begin{theorem}[Tipler]
Topology change cannot occur on an isochronous 4-geometry which
satisfies Einstein's equations and the weak energy condition.
\end{theorem}

Note that Einstein's equations, by themselves, place no constraint
whatsoever on the topology (or even the metric). It is only the
combination of the equations with some assumptions about physically
realistic energy momentum tensors that places a kinematic constraint.

\section{Manifolds which are not isochronous}
\label{sec:unorientable}
When time-orientability is not required, some timelike curves from
$\Sigma_1$ can enter the manifold and exit through $\Sigma_1$ rather
than $\Sigma_2$. A simple example is $S^1 \rightarrow \emptyset$, as
depicted in figure~\ref{fig:cap}, formed from the cylinder $S^1
\otimes I$, where I is an interval $[0,\tau$]. The first spacelike
surface, $\Sigma_1$, is $S^1$; in place of the second surface we
identify antipodal points to create a m\"{o}bius band oriented in the
space direction but not the time direction. Every timelike curve from
$\Sigma_1$ ($S^1$) which enters $M$ re-emerges through $\Sigma_1$
after a finite time.

Initial data on $\Sigma_1$ cannot be defined without knowing the value
of $\tau$ (the time at which the antipodal map occurs). Consider, for
example, a wave $\psi(\theta,t)$ which satisfies the one dimensional
wave equation. The general solution $\psi(\theta,t) = \zeta(\theta
-vt) + \phi(\theta +vt)$ (with $\zeta$ and $\phi$ having period
$2\pi$) is subject to additional constraints - consistency
requires: $\psi(\theta,t) = \zeta(\theta -vt) + \zeta(\theta +vt-
2v\tau+\pi)$, which depends upon the value of $\tau$.

Clearly the example can be extended to any number of disconnected
spaces $n \times S^1 \rightarrow m \times S^1 $. The model has been
described as the annihilation of a universe \cite{borde}. The main
purpose of this paper is to consider kinematics rather than dynamics,
however, this particular example is flat - and hence trivially
satisfies Einstein's equations in 1+1 dimensions. The model can also
be extended to higher dimensions. In 2 + 1 dimensions the space
orientation is also reversed by the antipodal map. In 3+1 dimensions
we have $S^3 \rightarrow \emptyset$ which satisfies Einstein's
equations with a non-zero cosmological constant (it is an Einstein
cylinder with the antipodal map imposed). The constraints on initial
data in the 3+1 case can be demonstrated by considering gravitational
waves following null geodesics rather than an unspecified extraneous
wave - in which case the manifold is the combination of an underlying
slowly varying spacetime and a smaller scale ripple.

If the topology change is associated with a measurement, then the
measurements at two different times, corresponding to a topology
change at $t=\tau_1$ or at $t=\tau_2$, would be incompatible. The
topology change cannot take place at both values of $\tau$, and the
alternative times would impose different boundary conditions on the
problem. To illustrate the non-Boolean logic; we consider either an
unspecified one dimensional wave, a classical object like a billiard
ball or (in 3+1 dimensions) a gravitational wave that is an intrinsic
part of the manifold. In either case, the initial conditions must
include a combination of a forward and backward moving wave (or
particle); and the combination depends upon the value of $\tau$. 

Consider a version of the two slit experiment; with slits $S_1$ and
$S_2$ and a measurement at $t= \tau$ to determine if a billiard ball
is in the region $X=[x_1,x_2]$. As described above we associate a
topology change and time reversal with the measurement process. The
experiment can be carried out with either $S_1$ or $S_2$ or both slits
open.

With the arrangement of figure~\ref{fig:slits}(a) with only slit $S_1$
open, there are no trajectories consistent with a measurement at $t =
\tau$. Similarly for the case with only $S_2$ open. With both $S_1$
and $S_2$ open, there is now a range of trajectories consistent with a
measurement at $t=\tau$ and detection at X - one of which is shown by
the dashed line in Figure~\ref{fig:slits}(b).

These examples show that the probability of going through $S_1$ and
reaching X is zero as is the probability of going through $S_2$ and
reaching X, but the probability of going through ($S_1$ or $S_2$) and
reaching X is not zero. This is a clear example of non-Boolean logic.

In a technical sense the objectives of this paper have been satisfied
by this simple, and well-known, example of a geometry which exhibits a
type of context sensitivity. Initial data cannot be specified
independently of the topology change; and there is a clear causal path
from the measurement (antipodal map) to the {\em initial} surface. The
parameter $\tau$ must be specified before the geometry can be defined.

Unfortunately, continuity prevents this construction being extended to
more realistic examples such as a connected boundary $\Sigma_1$
changing to a non-empty boundary $\Sigma_2$ with a different
topology. It is not therefore suitable as a model of particles
interacting {\em within} a universe.

\section{Spacetimes with closed timelike curves}
Considering compact isochronous geometries, topology change can occur
if CTC's are present in the interior of M. Some (or all) curves from
the initial surface $\Sigma_1$ have no endpoint, they are trapped near
a timelike curve. The final surface $\Sigma_2$ has some curves which
originated in $\Sigma_1$ and others that have no past endpoints
because they originated from the region of a CTC. There is no limit on
topology change in 4 dimensions if CTCs are allowed \cite{geroch}.

Trivial examples of topology change with CTCs are known
\cite{borde,dowker_garcia}; in which either the initial or final
surface is empty - (see for example figure~\ref{fig:s1CTC}).  There
are also examples with no timelike curve from the initial to final
surface, because they are all confined to the interior of M. In this
case $\Sigma_2$ cannot be considered to be in the future of
$\Sigma_1$. These cases will not be considered further because of the
lack of any causal connection between the two boundaries.

The examples of interest are those where curves from a subset $A
\subset \Sigma_1$ have an endpoint in $B \subset \Sigma_2$ while
others are trapped within $M$. Since $\Sigma_2$ is compact there must
be at least one point, $p$, of $\Sigma_2$ which is not in $B$. The
curve through $p$ cannot have a past endpoint in $M$ and must
therefore originate from the region of a CTC.

These geometries do not exhibit quantum phenomena. The topology
change introduces unknown (and probably unknowable) boundary
conditions because some timelike curves do not originate in the
initial surface. However, the topology change places no constraint on
the initial data; there is no causal link from the measurement to the
initial state nor any clear mechanism for non-local effects between
separated measurements, as seen in quantum mechanics. 

Although there is no dynamical constraint on the formation of CTCs,
Tipler's theorem shows that Einstein's equations, plus the weak energy
condition, prevents this mechanism for topology change.

\section{Non-compact spacetimes}
Geroch's theorem can be applied to externally simple geometries, even
when they are not compact, because the topology change takes place in
a compact region. An externally simple spacetime would be an
appropriate description for a geon embedded in $\Rset^4$. However,
continuity still prevents a failure of time-orientability from being a
mechanism for topology change.

More generally topology change can take place in non-compact
geometries\cite{krasnikov94}, because the timelike curves from
$\Sigma_1$ can avoid the boundaries without being trapped near a
closed timelike curve - see for example figure~\ref{fig:s12r1}. These
counter-examples can have topology change without a breakdown of the
causal structure.

\section{Non-spacelike boundaries}

A further relaxation of the assumptions in Geroch's theorem, is to
have boundaries (one or both) that are not entirely spacelike. The
terms {\em initial} and {\em final} for the boundaries are not
strictly correct although the majority of the surface could be
spacelike. With such a significant departure from the conditions of
Geroch's theorem, topology change is easily
demonstrated. Figures~\ref{fig:trousers1} and \ref{fig:trousers2} show
a topology change from $S^1 \rightarrow S^1 \cup S^1$, both examples
are not time-orientable. Figure~\ref{fig:trousers1} has an initial
spacelike boundary, but the final boundaries are not
spacelike. Figure~\ref{fig:trousers2} does not have either an initial
or final spacelike boundary but it is clearly flat, and therefore a
trivial solution to the field equations. Another feature of
figure~~\ref{fig:trousers2} is the causal link between the two {\em
legs} of the trousers, which has some comparisons with an EPR
experiment. A more realistic model of an EPR experiment would be a
combination of three topology changing regions corresponding to the
pair creation, and measurement in the two arms of the experiment.

An externally simple example is shown in figure~\ref{fig:droplet}
which illustrates $\Rset \rightarrow \Rset \cup S^1$: the manifold is
not isochronous. The mechanism for topology change is essentially
that of the non-isochronous case (section~\ref{sec:unorientable}
above), the freedom to allow non-spacelike surfaces, permits the
construction of non-trivial examples with a continuous metric. In the
examples of figures~\ref{fig:trousers1}~and~\ref{fig:droplet}, the
initial surface $\Sigma_1$ is spacelike and time {\em can} be oriented
in a finite neighbourhood of $\Sigma_1$. It is the subsequent topology
change that turns some timelike geodesics back to $\Sigma_1$ and
obstructs the time-orientability of the geometry.

\section{Conclusion}
In these examples, boundary conditions cannot be specified on the
initial surface without knowing about the subsequent measurement
(topology change). There is a causal link between the measurement and
the initial surface. This breakdown of the causal structure,
associated with a measurement, will necessarily require a context
dependent theory such as quantum theory to describe the observations.

All these compact and externally simple examples prevent, or at least
limit, the construction of spacelike boundaries and spacelike
hypersurfaces. It must be stressed that this does not indicate any
departure from the equations or structure of classical general
relativity. To require spacelike boundaries is an extra, convenient
but unwarranted, constraint commonly imposed upon the theory. It is a
constraint that is not justified either by the mathematical structure
or by physical reality. Indeed, what we know about quantum theory is
incompatible with spacetime being a classical hypersurface evolving
with time. Quantum theory gives evidence for the small scale structure
of spacetime and general relativity could, in return, explain the
origins of quantum phenomena.

Of all the interpretations of quantum mechanics, this work relates
closely to the pragmatic interpretation of quantum mechanics, where
the wavefunction is simply a way of determining the probabilities of
the different outcomes from an experiment.

Consider the classical case, $P(n)= 1/6; (n=1,2\dots 6)$ is a
probability function which gives the result of a dice throw; but there
is no direct relationship between $P(n)$ and the underlying Newtonian
laws of motion. Indeed the probability function does not depend upon
the exact form of the equations of motion, it depends only upon the
structural form and the symmetry. The classical equations can be
expressed as deterministic equations with uniquely defined
trajectories determined by the initial conditions alone. The sets of
initial conditions corresponding to different results satisfy a
Boolean algebra. In principle a measure on the space of initial
conditions can be used to calculate the probability function,
$P(n)$. The equations of motion are independent of $n$ - this gives
the essential symmetry.

With the measurement dependent effects shown here, general relativity
gives a different structure. A measure over {\em initial} conditions
and measurement conditions would also give a Boolean logic as required
by Mackey \cite{mackey}; but a measure over {\em initial} conditions
alone would give a different logical structure which must be
orthomodular\cite{hadley97,beltrametti_cassinelli}. The orthomodular
structure can be represented by the projections on a Hilbert Space
\cite{beltrametti_cassinelli}. Given the Hilbert Space structure, it
is well-known that the equations for the wavefunction can be derived
from symmetry arguments\cite{ballentine,weinberg95}. As in the
classical case, the probability function does not depend upon the
details of the equations of motion; just on their form.

\newcommand{\trousers}{
    \linethickness{0.5mm}
    \renewcommand{\yscale}{0.25}
    \scaleput(40,60){\arc(-30,0){180}}
    \scaleput(15,360){\bigcircle{30}}
    \scaleput(65,360){\bigcircle{30}}

    \curvedashes{0,1,2}
    \scaleput(40,60){\arc(-30,0){-180}}
    \renewcommand{\yscale}{1.0}
    \curvedashes{}

    \linethickness{0.2mm}
    \put(72,15){\makebox(0,0)[bl]{$\Sigma_1$}}
    \put(76,93){\makebox(0,0)[bl]{$\Sigma_2$}}
    \put(4,93){\makebox(0,0)[br]{$\Sigma_2$}}
}

\newpage
\setlength{\unitlength}{0.5mm}
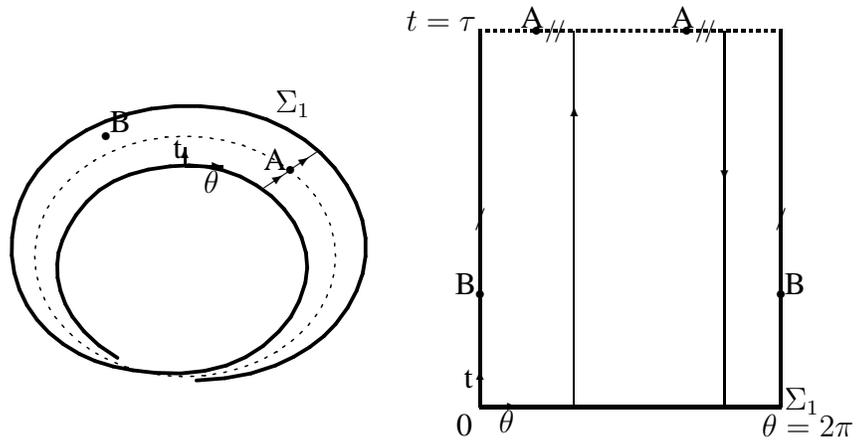
\begin{figure}[h]
\center{
\begin{tabular}{ll}
\begin{picture}(90,220)(-10,-10)
\linethickness{0.5mm}

\curve(53,7,74,11,90,22,98,40,93,59,77,74,54,80,29,76,11,63,4,45,8,27,22,14,42,9,61,
11,75,20,82,33,81,45,73,56,60,63,46,64,31,60,21,51,16,38,20,24,32,13)

\linethickness{0.2mm}
\curvedashes{0,1,2}
\renewcommand{\yscale}{0.8}
\scaleput(50,50){\bigcircle{80}}
\curvedashes{}
\renewcommand{\yscale}{1.0}

\put(29,72){\circle*{2}}
\put(30,73){\makebox(0,0)[bl]{B}}
\put(78,63){\circle*{2}}
\put(77,63){\makebox(0,0)[br]{A}}

\put(74,77){\makebox(0,0)[bl]{$\Sigma_1$}}

\linethickness{0.2mm}
\put(50,64){\vector(0,1){5}}
\put(50,64){\vector(1,0){10}}
\put(48,66){\makebox(0,0)[b]{t}}
\put(55,60){\makebox(0,0)[l]{$\theta$}}

\put(71,58.333){\line(3,2){14}}
\put(71,58.333){\vector(3,2){5}}
\put(78,63){\vector(3,2){5}}
\end{picture}&

\begin{picture}(90,220)(-40,-10)
\linethickness{0.5mm}
\put(0,0){\line(0,1){100}}
\put(0,0){\line(1,0){80}}
\put(80,0){\line(0,1){100}}
\put(0,100){\dashbox{1}(80,0){}}

\put(0,50){\makebox(0,0){/}}
\put(80,50){\makebox(0,0){/}}
\put(20,100){\makebox(0,0){//}}
\put(60,100){\makebox(0,0){//}}

\put(0,30){\circle*{2}}
\put(-1,30){\makebox(0,0)[br]{B}}
\put(80,30){\circle*{2}}
\put(81,30){\makebox(0,0)[bl]{B}}

\put(15,100){\circle*{2}}
\put(17,101){\makebox(0,0)[br]{A}}
\put(55,100){\circle*{2}}
\put(57,101){\makebox(0,0)[br]{A}}

\put(81,-2){\makebox(0,0)[bl]{$\Sigma_1$}}

\linethickness{0.2mm}
\put(-1,100){\makebox(0,0)[br]{$t=\tau$}}
\put(75,-5){\makebox(0,0)[l]{$\theta = 2 \pi$}}

\put(0,0){\vector(0,1){10}}
\put(0,0){\vector(1,0){10}}
\put(-3,5){\makebox(0,0)[b]{t}}
\put(5,-4){\makebox(0,0)[l]{$\theta$}}
\put(-2,-2){\makebox(0,0)[tr]{0}}

\put(25,0){\line(0,1){100}}
\put(65,0){\line(0,1){100}}
\put(25,70){\vector(0,1){10}}
\put(65,70){\vector(0,-1){10}}

\end{picture}
\end{tabular}
}
\caption{Topology change from $S^1\rightarrow \emptyset$ to create a
manifold that is not isochronous. Depicted pictorially and
diagrammatically as a rectangle with sides identified. Every timelike
curve originating in $\Sigma_1$ returns to $\Sigma_1$}
\label{fig:cap}
\end{figure}

\newpage
\setlength{\unitlength}{0.5mm}
\begin{figure}[h]
\center{
\begin{tabular}{ll}

\begin{picture}(110,220)(-40,-10)
\linethickness{0.5mm}
\put(0,0){\line(0,1){100}}
\put(0,0){\line(1,0){80}}
\put(80,0){\line(0,1){100}}
\put(0,100){\dashbox{1}(80,0){}}

\put(0,50){\makebox(0,0){/}}
\put(80,50){\makebox(0,0){/}}

\put(81,-2){\makebox(0,0)[bl]{$\Sigma_1$}}

\linethickness{0.2mm}
\put(-1,100){\makebox(0,0)[br]{$t=\tau$}}
\put(75,-5){\makebox(0,0)[l]{$\theta = 2 \pi$}}

\put(0,0){\vector(0,1){10}}
\put(0,0){\vector(1,0){10}}
\put(-3,5){\makebox(0,0)[b]{t}}
\put(5,-4){\makebox(0,0)[l]{$\theta$}}
\put(-2,-2){\makebox(0,0)[tr]{0}}

\linethickness{1mm}
\put(15,100){\line(1,0){5}}
\put(17,106){\makebox(0,0){X}}
\put(55,100){\line(1,0){5}}
\put(57,106){\makebox(0,0){X}}

\put(0,95){\line(1,0){17}}
\put(19,93){\makebox(0,0)[tr]{$S_1$}}
\put(25,95){\line(1,0){55}}

\end{picture}& 

\begin{picture}(110,220)(-40,-10)
\linethickness{0.5mm}
\put(0,0){\line(0,1){100}}
\put(0,0){\line(1,0){80}}
\put(80,0){\line(0,1){100}}
\put(0,100){\dashbox{1}(80,0){}}

\put(0,50){\makebox(0,0){/}}
\put(80,50){\makebox(0,0){/}}

\put(81,-2){\makebox(0,0)[bl]{$\Sigma_1$}}

\linethickness{0.2mm}
\put(-1,100){\makebox(0,0)[br]{$t=\tau$}}
\put(75,-5){\makebox(0,0)[l]{$\theta = 2 \pi$}}

\put(0,0){\vector(0,1){10}}
\put(0,0){\vector(1,0){10}}
\put(-3,5){\makebox(0,0)[b]{t}}
\put(5,-4){\makebox(0,0)[l]{$\theta$}}
\put(-2,-2){\makebox(0,0)[tr]{0}}
\linethickness{1mm}
\put(15,100){\line(1,0){5}}
\put(17,106){\makebox(0,0){X}}
\put(55,100){\line(1,0){5}}
\put(57,106){\makebox(0,0){X}}

\put(0,95){\line(1,0){17}}
\put(19,93){\makebox(0,0)[tr]{$S_1$}}
\put(25,95){\line(1,0){32}}
\put(59,93){\makebox(0,0)[tr]{$S_2$}}
\put(65,95){\line(1,0){15}}

\linethickness{0.1mm}
\curvedashes{0,3,2}
\curve(10,0, 18,100)
\curve(66,0, 58,100)

\end{picture}
\end{tabular}
}
\caption{A Two slit experiment: a) there are no trajectories consistent with only slit 1 or slit 2 being open, b) there are consistent trajectories (dashed line) when both slits are open.}
\label{fig:slits}
\end{figure}
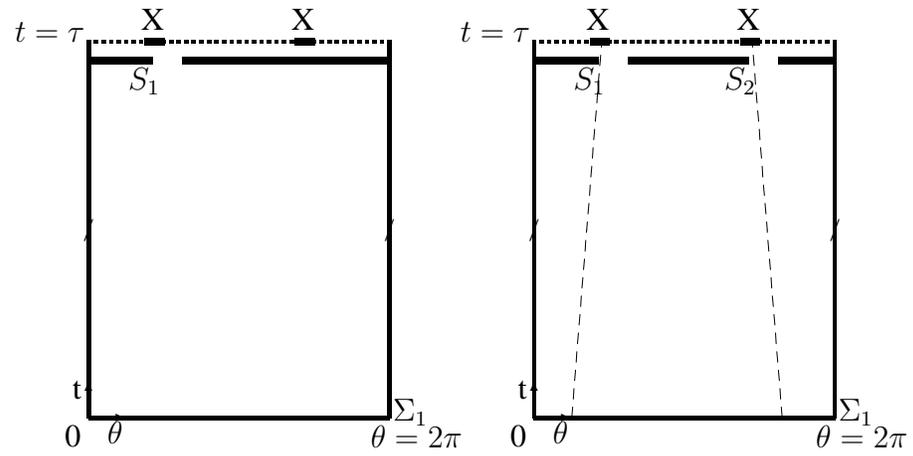

\newpage
\setlength{\unitlength}{0.5mm}
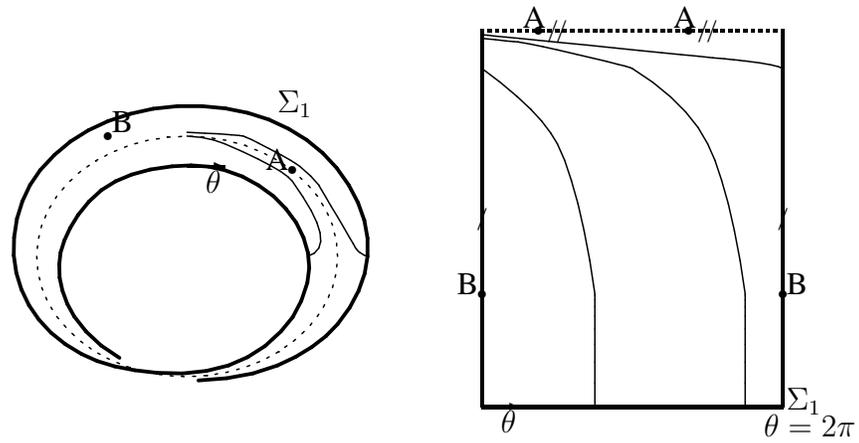
\begin{figure}[h]
\center{
\begin{tabular}{ll}
\begin{picture}(90,220)(-10,-10)
\linethickness{0.5mm}
\curve(53,7,74,11,90,22,98,40,93,59,77,74,54,80,29,76,11,63,4,45,8,27,22,14,42,9,61,
11,75,20,82,33,81,45,73,56,60,63,46,64,31,60,21,51,16,38,20,24,32,13)

\linethickness{0.2mm}
\tagcurve(100,40, 98,40, 96,41, 95,42, 90,50.5, 85,59,  80,64, 75,67, 65,72, 50,73, 40,73)
\tagcurve(10,40, 82,40, 78,60, 60,70, 50,72, 10,72)

\curvedashes{0,1,2}
\renewcommand{\yscale}{0.8}
\scaleput(50,50){\bigcircle{80}}
\curvedashes{}
\renewcommand{\yscale}{1.0}

\put(29,72){\circle*{2}}
\put(30,73){\makebox(0,0)[bl]{B}}
\put(78,63){\circle*{2}}
\put(77,63){\makebox(0,0)[br]{A}}

\put(74,77){\makebox(0,0)[bl]{$\Sigma_1$}}

\linethickness{0.2mm}
\put(50,64){\vector(1,0){10}}
\put(55,60){\makebox(0,0)[l]{$\theta$}}
\end{picture}&

\begin{picture}(90,220)(-40,-10)
\linethickness{0.5mm}
\put(0,0){\line(0,1){100}}
\put(0,0){\line(1,0){80}}
\put(80,0){\line(0,1){100}}
\put(0,100){\dashbox{1}(80,0){}}

\put(0,50){\makebox(0,0){/}}
\put(80,50){\makebox(0,0){/}}
\put(20,100){\makebox(0,0){//}}
\put(60,100){\makebox(0,0){//}}

\put(0,30){\circle*{2}}
\put(-1,30){\makebox(0,0)[br]{B}}
\put(80,30){\circle*{2}}
\put(81,30){\makebox(0,0)[bl]{B}}

\put(15,100){\circle*{2}}
\put(17,101){\makebox(0,0)[br]{A}}
\put(55,100){\circle*{2}}
\put(57,101){\makebox(0,0)[br]{A}}

\put(81,-2){\makebox(0,0)[bl]{$\Sigma_1$}}

\linethickness{0.2mm}
\put(75,-5){\makebox(0,0)[l]{$\theta = 2 \pi$}}
\tagcurve(30,-20, 30,0, 30,30, 20,70, 0,90, -10,92)
\tagcurve(100,70, 80,90, 70,92,  20,97, 10,98, 0,99, -10,99.5)

\tagcurve(70,-20, 70,0, 70,30, 60,70, 40,90, 20,95, 10,97, 0,98, -10,99)
\put(0,0){\vector(1,0){10}}
\put(5,-4){\makebox(0,0)[l]{$\theta$}}

\end{picture}
\end{tabular}
}
\caption{Topology change $S^1 \rightarrow \emptyset$ on a manifold
with CTCs. The timelike curves form $\Sigma_1$ enter the manifold and
never exit. They asymptotically approach the CTC (dashed line).}
\label{fig:s1CTC}
\end{figure}

\newpage
\setlength{\unitlength}{1mm}
\begin{figure}[h]
\center{
\begin{picture}(90,110)(0,-10)
\linethickness{0.5mm} 
\renewcommand{\yscale}{0.5} 
\curve(105,102, 59.5,103.9, 53.4,105.9, 50.8,107.7, 49.0,109.5,
47.5,111.2, 41.1,116.8, 33.5,119.7, 25.5,119.4, 18.2,116.1,
12.8,110.3, 10.2,102.8, 10.6,94.8, 14.1,87.7, 20.2,82.5, 27.8,80.1,
35.8,80.8, 43.1,84.5, 49.3, 90.7, 51.2,92.5, 54.1,94.4, 61.5,96.3,
105,98)
\scaleput(30,30){\bigcircle{40}}
\put(9,16){\makebox(0,0)[br]{$\Sigma_1$}}
\put(9,51){\makebox(0,0)[br]{$\Sigma_2$}} 
\renewcommand{\yscale}{1}
\linethickness{0.2mm}
\put(10,15){\vector(0,1){34}}
\put(30,5){\vector(0,1){35}}
\put(35,6){\vector(1,4){9}}
\put(40,6.5){\vector(1,2){20.5}}
\put(50,13){\vector(3,2){54}}
\end{picture}
}\caption{Topology change $S^1 \rightarrow \Rset$ on a manifold that is not compact. There is no breakdown of the causal structure.}
\label{fig:s12r1}
\end{figure}
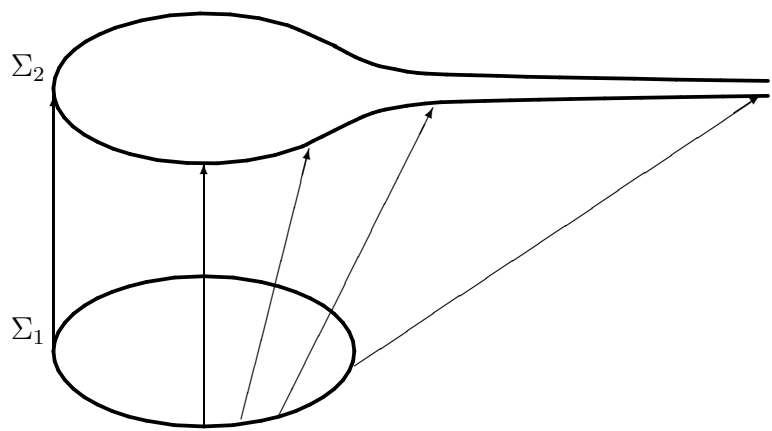

\newpage
\setlength{\unitlength}{0.5mm}
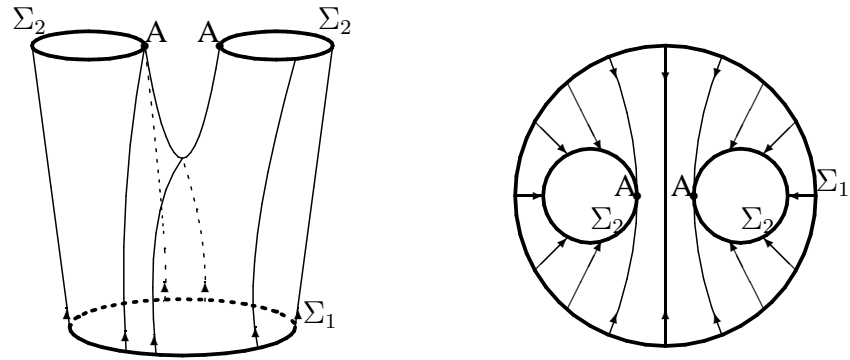
\begin{figure}[h]
\center{
\begin{tabular}{ll}
\begin{picture}(90,220)(-10,-10)
\linethickness{0.5mm}
\trousers
    \put(30,91){\makebox(0,0)[bl]{A}}
    \put(30,90){\circle*{2}}
    \put(50,91){\makebox(0,0)[br]{A}}
    \put(50,90){\circle*{2}}
       
\linethickness{0.2mm}
\curve(30,90, 40,60, 50,90)

 \curve(33,8, 33,40, 40,60)
\put(32.5,13){\vector(0,1){0.1}}
\curve(25,9, 25,50, 30,90)
\put(24.8,14){\vector(0,1){0.1}}
\curve(10,15, 0,90)
\put(9.25,20){\vector(0,1){0.1}}
\curve(70,15, 80,90)
\put(70.75,20){\vector(0,1){0.1}}
\curve(60,10, 60,40, 70,86)
\put(59.5,15){\vector(0,1){0.1}}
\curvedashes{0,1,2}
  \curve(35,22, 35,40, 30,90)
  \put(35.3,27){\vector(0,1){0.1}}
  \curve(46,22, 45,40, 40,60)
  \put(46,27){\vector(0,1){0.1}}
\curvedashes{}

\end{picture}&

\begin{picture}(90,220)(-40,-10)
\linethickness{0.5mm}
\put(40,50){\bigcircle{80}}
\put(20,50){\bigcircle{25}}
\put(60,50){\bigcircle{25}}

\put(80,50){\makebox(0,0)[bl]{$\Sigma_1$}}
\put(20,40){\makebox(0,0)[bl]{$\Sigma_2$}}
\put(60,40){\makebox(0,0)[bl]{$\Sigma_2$}}

\linethickness{0.2mm}
\put(32.5,50){\circle*{2}}
\put(32.5,50){\makebox(0,0)[br]{A}}
\put(47.5,50){\circle*{2}}
\put(47.5,50){\makebox(0,0)[br]{A}}

\put(0,50){\vector(1,0){7.5}}
\put(80,50){\vector(-1,0){7.5}}

\put(40,10){\line(0,1){80}}
\curve(25,13, 32.5,50, 25,87)
\curve(55,13, 47.5,50, 55,87)
\put(25,13){\vector(1,3){2}}
\put(55,13){\vector(-1,3){2}}
\put(55,87){\vector(-1,-3){2}}
\put(25,87){\vector(1,-3){2}}

\put(40,10){\vector(0,1){10}}
\put(40,90){\vector(0,-1){10}}

\put(5,30){\vector(1,1){9}}
\put(75,30){\vector(-1,1){9}}
\put(75,70){\vector(-1,-1){9}}
\put(5,70){\vector(1,-1){9}}

\put(14,20){\vector(1,2){9}}
\put(66,20){\vector(-1,2){9}}
\put(66,80){\vector(-1,-2){9}}
\put(14,80){\vector(1,-2){9}}

\end{picture}
\end{tabular}
}
\caption{Topology change from $S^1\rightarrow S^1 \cup
S^1$. $\Sigma_1$ is spacelike but some timelike curves from
$\Sigma_1$return to $\Sigma_1$. The surface $\Sigma_2$ is not entirely
spacelike. At points $A$ the timelike curves are tangent to
$\Sigma_2$. }
\label{fig:trousers1}
\end{figure}

\newpage

\setlength{\unitlength}{0.5mm}
\begin{figure}[h]
\center{
\begin{tabular}{ll}
\begin{picture}(90,220)(-10,-10)
\linethickness{0.5mm}
\trousers
    \put(40,7){\makebox(0,0)[tl]{A}}
    \put(40,7.5){\circle*{2}}
    \put(40,22){\makebox(0,0)[tl]{A}}
    \put(40,22.5){\circle*{2}}
    \put(15,87){\makebox(0,0)[bl]{A}}
    \put(15,86.25){\circle*{2}}
    \put(15,95){\makebox(0,0)[bl]{A}}
    \put(15,93.75){\circle*{2}}
    \put(65,87){\makebox(0,0)[bl]{A}}
    \put(65,86.25){\circle*{2}}
    \put(65,95){\makebox(0,0)[bl]{A}}
    \put(65,93.75){\circle*{2}}
    
\linethickness{0.2mm}
\curve(30,8, 40,15, 50,8)
\curve(25,9, 40,30, 55,9)
  \tagcurve(13,10, 13,12, 10,70, 15,86.25, 24.5,70)
  \tagcurve( 55.5,70,  65,86.25, 70,70, 67,12, 67,10)
\curve(10,15, 0,90)
\put(9.25,20){\vector(0,1){0.1}}
\curve(70,15, 80,90)
\put(70.75,20){\vector(0,1){0.1}}
  \curve(20,10, 18,45, 20,50, 29,45)
  \curve(29,45, 40,40, 51,45)
  \curve(51,45,  60,50, 62,45, 60,10)

\curvedashes{0,1,2}
  \curve(30,90, 40,60, 50,90)
  \curve(15,86.25, 17,84, 24.5,70, 40,55, 55.5,70, 63,84, 65,86.25)
\curvedashes{}

\end{picture}&

\begin{picture}(90,220)(-40,-10)
\linethickness{0.5mm}
\put(40,50){\bigcircle{80}}
\put(20,50){\bigcircle{25}}
\put(60,50){\bigcircle{25}}

\put(80,50){\makebox(0,0)[bl]{$\Sigma_1$}}
\put(25,50){\makebox(0,0)[b]{$\Sigma_2$}}
\put(65,50){\makebox(0,0)[b]{$\Sigma_2$}}

\linethickness{0.2mm}
\put(20,62.5){\circle*{2}}
\put(20,63.5){\makebox(0,0)[b]{A}}
\put(20,37.5){\circle*{2}}
\put(20,38.5){\makebox(0,0)[b]{A}}

\put(60,62.5){\circle*{2}}
\put(60,63.5){\makebox(0,0)[b]{A}}
\put(60,37.5){\circle*{2}}
\put(60,38.5){\makebox(0,0)[b]{A}}

\put(40,90){\circle*{2}}
\put(40,91){\makebox(0,0)[b]{A}}
\put(40,10){\circle*{2}}
\put(40,9){\makebox(0,0)[t]{A}}

\put(0,50){\vector(1,0){7.5}}
\put(80,50){\vector(-1,0){7.5}}

\put(2.5,62.5){\line(1,0){17.5}}
\put(2.5,37.5){\line(1,0){17.5}}
\put(60,62.5){\line(1,0){17.5}}
\put(60,37.5){\line(1,0){17.5}}
\put(9,75){\line(1,0){62}}
\put(9,25){\line(1,0){62}}
\put(20,84){\line(1,0){40}}
\put(20,16){\line(1,0){40}}

\curvedashes{0,1,2}
  \curve(32.5,50, 47.5,50)
  \curve(20,37.5, 60,37.5)
  \curve(20,62.5, 60,62.5)
\curvedashes{}

\end{picture}
\end{tabular}
}
\caption{Topology change from $S^1\rightarrow S^1 \cup S^1$. The
manifold is flat but neither $\Sigma_1$ nor $\Sigma_2$ is entirely
spacelike. At points $A$ the timelike curves are tangent to the surfaces.
 Note the timelike curves (dashed curves) joining the
different legs of the trousers.}
\label{fig:trousers2}
\end{figure}
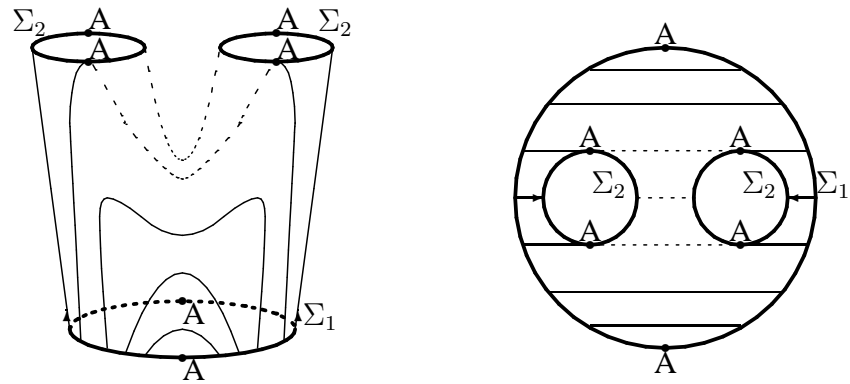

\newpage
\setlength{\unitlength}{0.7mm}
\begin{figure}[h]
\center{
\begin{picture}(100,120)(0,-10)
\linethickness{0.5mm}
\put(0,10){\line(1,0){100}}
\put(0,60){\line(1,0){100}}
\put(50,50){\bigcircle{10}}
\put(0,0){\makebox(0,0)[bl]{$\Sigma_1 = \Rset$}}
\put(0,62){\makebox(0,0)[bl]{$\Sigma_2 = \Rset \cup S^1$}}

\linethickness{0.2mm}
\put(10,10){\vector(0,1){50}}
\put(20,10){\vector(0,1){50}}
\put(30,10){\vector(0,1){50}}
\put(50,10){\vector(0,1){35}}
\put(70,10){\vector(0,1){50}}
\put(80,10){\vector(0,1){50}}
\put(90,10){\vector(0,1){50}}

\tagcurve(35,0, 35,10, 37,45, 50,60,  63,45,   65,10, 65,0)
\tagcurve(40,0, 40,10, 42,45, 50,55,  58,45,   60,10, 60,0)
\tagcurve(45,0, 45,10, 46,47, 50,38)
\tagcurve(55,0, 55,10, 54,47, 50,38)

\put(50,60){\circle*{2}}
\put(50,61){\makebox(0,0)[br]{A}}
\put(50,55){\circle*{2}}
\put(50,50){\makebox(0,0)[b]{A}}

\end{picture}}
\caption{Topology change from $\Rset \rightarrow S^1 \cup
\Rset$. $\Sigma_1$ is spacelike but some timelike curves from
$\Sigma_1$return to $\Sigma_1$. The surface $\Sigma_2$ is not entirely
spacelike. At points $A$ the timelike curves are tangent to
$\Sigma_2$.}
\label{fig:droplet}
\end{figure}

\end{document}